# Towards a Connected Heterogeneous All-Medium Integrated Network (CHAIN) for Converged Connectivity Across Land, Sea, Air, and Space

P. A. Haigh, S. Rajbhandari, K. Bottrill, A. Trichili, S. Watson, H. Abumarshoud, D. Benton, S. Sinanovic, J. Herrnsdorf, P. Christopher, M. Khalily, R. Tafazolli, H. Haas, M. Lavery and W. Popoola

*Abstract*—**Next-generation connectivity depends on data traversing multiple physical media within a single end-to-end path, yet research in optical fibre, free-space optical and radio wireless, non-terrestrial networks, and underwater communications has advanced largely in isolation. This fragmentation has a wider adverse impact on communication performance, deployment and adaption as the most demanding open problems in next-generation connectivity, cross-medium channel characterisation, transport-layer protocol design across heterogeneous latency regimes, and cross-domain orchestration, sit at the boundaries between domains and can only be addressed by bringing these challenges together in cross-domain systems. Space-air-ground integrated network research has begun to treat three domains analytically, and its recent extension to the sea surface represents the most ambitious multi-domain framework proposed to date, yet neither has produced experimental results, a deployed system, or a standards engagement. The subsurface and maritime domain remains absent from both. This paper proposes the connected heterogeneous all-medium integrated network (CHAIN), a framework that treats the undersea, maritime, terrestrial, aerial and space domains as a single design space. We identify an all-photonic network backbone as the unifying physical infrastructure, Artificial intelligence (AI)-driven orchestration as the cross-domain control layer, and the joints between domains as a unification challenge. We survey the state of the art across all domains and the existing cross-domain literature, develop the CHAIN framework and its technical pillars, characterise the five core open problems the field must resolve, and set out a research roadmap from near-term measurement campaigns and testbed validation through cross-medium field trials to global-scale deployment.**

## I. INTRODUCTION

GLOBAL connectivity demands increasingly require data to traverse multiple physical media within a single end-to-end path. Essential services such as climate monitoring, disaster response coordination, offshore energy infrastructure, precision agriculture and global logistics all depend on communication links that cross between undersea, maritime, terrestrial, aerial and space domains, often within latency, capacity and reliability constraints that no single transmission technology can satisfy. Consider the future offshore wind farm operation where structural sensor data is exchanged via underwater optical wireless communication (UOWC) links and subsea fibre to an onshore substation; LiFi delivers high-bandwidth condition-monitoring and control links inside the nacelle and substation without causing radio frequency (RF) interference in equipment-dense enclosures; routine inspection is handled by drones relaying high-bandwidth imagery over free-space optics (FSO) and/or RF links; satellite backhaul covers areas beyond conventional wireless reach; and grid integration data flows through terrestrial fibre to the national network, illustrated in Fig. 1. Wind farms of this scale already operate globally, with exactly this connectivity problem, and no unified network architecture exists to address it.

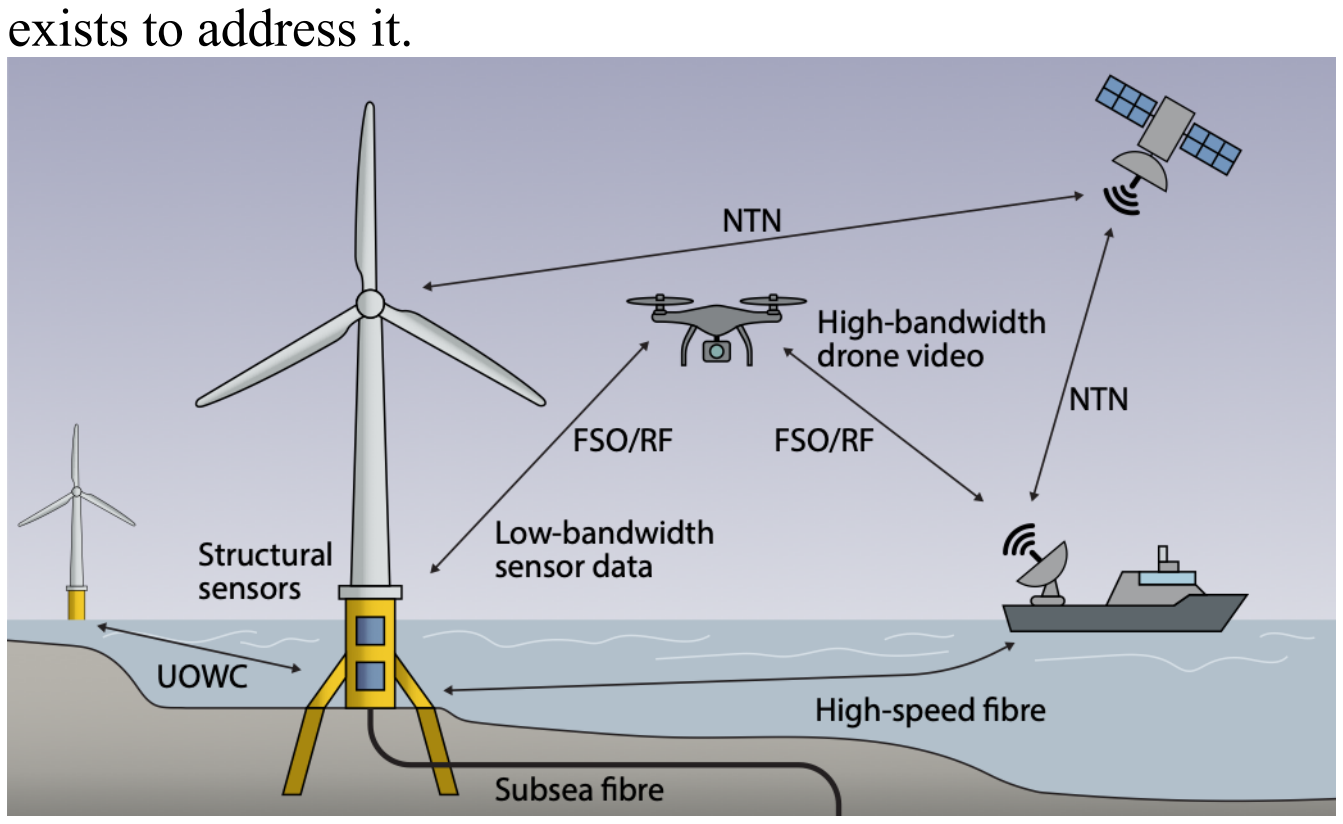

**Fig.1: An example of a modern scenario where multi-medium transmission is required to operate seamlessly across interfaces.**

The constituent technologies have each advanced considerably, but within largely self-contained communities with limited cross-domain dialogue. Optical fibre capacity has reached the petabit-per-second scale in laboratory conditions, most recently with NICT

P. A. Haigh is with Centre for Networks, Communications and Systems, School of Electronic Engineering and Computer Science, Queen Mary University of London, London, E1 4NS, UK

S. Rajbhandari and J. Herrnsdorf are with the Institute of Photonics, University of Strathclyde, Glasgow, UK.

K. Bottrill is with Optoelectronics Research Centre, University of Southampton, UK.

A. Trichili is with the School of Engineering, Physics and Mathematics, Northumbria University, Newcastle upon Tyne NE1 8ST

S. Watson, H. Abumarshoud and M. Lavery are with the University of Glasgow, Glasgow, UK.

D. Benton is with Aston Institute of Photonic Technologies, Aston University, Birmingham B4 7ET, UK.

S. Sinanovic is with the School of Engineering and Built Environment, Glasgow Caledonian University, Glasgow G4 0BA, UK

. Christopher is with Department of Electrical and Electronic Engineering, University of Nottingham, Nottingham, UK.

M. Khalily and R. Tafazolli are with the Institute for Communication Systems (ICS), 5G/6G Innovation Centre, University of Surrey, Guildford, GU2 7XH, UK.

H. Haas is with LiFi Research and Development Centre, Department of Engineering, University of Cambridge, Cambridge, UK

W. Popoola is with The Institute for Imaging, Data and Communications, School of Engineering, University of Edinburgh, Edinburgh, EH9 3FG, UK.

and Sumitomo Electric recording 1.02 Pb/s over 1,808 km using 19-core multicore fibre [1, 2], while deployed transoceanic systems continue to scale through coherent detection and digital signal processing improvements [3-6]. Satellite connectivity has matured rapidly: 3rd Generation Partnership Project (3GPP) Release 17 introduced the first normative specifications for low earth orbit (LEO) and geostationary earth orbit (GEO) satellites and implicit support for high-altitude platform systems (HAPS) in the 5G New Radio (NR) and Internet-of-things (IoT)-non-terrestrial network (NTN) air interface, and Release 18 extended this to Ka-band operation and improved mobility management [7], and Release 19 has added regenerative payload functionality [8]. The scope of that standardisation work stops at the terrestrial-to-space interface and does not address interoperability with optical or underwater domains. FSO and RF wireless links have been characterised across a wide range of propagation environments, and hybrid RF/FSO architectures are well established as a means of maintaining link availability under various atmospheric conditions [9]. Underwater optical and acoustic communications, less visible in the mainstream broad telecommunications literature, have advanced considerably for ocean science and offshore energy applications, with optical links demonstrated at data rates approaching those of short-range terrestrial wireless in clear water [10]. Meanwhile, maritime links, which face intensified atmospheric challenges alongside extreme pointing, acquisition and tracking (PAT) requirements, have seen real-world demonstrations largely restricted to the defence context [11]. LiFi has also progressed substantially, with laboratory demonstrations reporting data rates in the Gb/s range and theoretical studies promising aggregate rates in the Tb/s range [12], while the IEEE 802.11bb standard represents a key milestone in its commercialisation.

Cross-domain integration has attracted growing attention, concentrated mainly on two-domain pairings. Satellite-terrestrial network integration has been studied extensively within the 5G and early 6G communities, addressing handover protocols, interference management and hybrid RF architectures [13]. Relay-assisted RF/FSO systems have been proposed to extend FSO reach through RF fallback under atmospheric disruption [14]. Unmanned aerial vehicle (UAV)-assisted relay architectures have been investigated for coverage extension and rapid emergency communications [15], and recently, reports address underwater-to-surface links combining technologies for shallow-water sensor networks [16-18]. In each case, the integration work typically takes two domains and holds the rest of the network as either terrestrial or fixed. Interoperability and standardisation gaps compound across every domain boundary in the path, and any system that crosses more than two of them runs into interface problems that no existing standard resolves.

Including each domain as a simultaneous design problem is a different undertaking, and one the literature has not yet addressed. How signal impairments accumulate across transitions between optical-underwater, FSO, RF and fibre segments within a single end-to-end path remains an open modelling question. Transport-layer protocols, designed for broadly homogeneous links, do not have established behaviour across the orders-of-magnitude latency difference between fibre and deep acoustic channels. The orchestration problem, routing across media with radically different availability, capacity and failure profiles in real time, has no solution that applies across more than two domains. Standards fragmentation reinforces all the above: 3GPP, ITU-R, ITU-T and the fragmented underwater communications standards landscape, spanning NATO STANAG 4748, ISO TC43/SC3 and IEEE OES, each address their own domain with no mechanisms for the connections between them, illustrated conceptually in Fig. 2.

**Fig. 2: Disciplinary silos are a problem for cross-pollination; cross-medium research has been traditionally performed in pairs (RF-FSO, fibre-RF, fibre-FSO), however, future generations will require substantial cross-cutting research to maximise link capacity and robustness.**

Table 1: A comparison of various connectivity domains

| Domain | Typical data rate (deployed) | Typical one-way latency | Operational range | Primary standards body | Domain TRL | Cross-medium integration TRL | Headline integration open problem |
|---|---|---|---|---|---|---|---|
| Optical fibre | Multi-Tb/s per fibre pair | ~5 µs/km | Global | ITU-T, IEEE 802.3 | 9 | 2-3 | Photonic interface to wireless and acoustic domains |
| FSO | 100 Mb/s-10 Gb/s | µs-ms | Metres to ~20 km | ITU-R, 3GPP, IEEE 802.11 | 8-9 | 3-4 | Cross-medium channel modelling at air-water and air-fibre interfaces |
| RF | kb/s to ~20 Gb/s peak (5G mmWave) | <1 ms to multi-second (ELF) | Local to global | 3GPP, ITU-R, IEEE 802.11 | 9 | 4-5 | Programmable RF interfaces and cross-domain protocols across IoT-NTN |
| NTN | 25-400 Mb/s (user terminal); 80-100 Gb/s (per LEO satellite) | 20-40 ms (LEO); <1 ms (HAPS) | Global (LEO); regional (HAPS) | 3GPP (Rel. 17/18), ITU-R | 8-9 (LEO); 4-5 (HAPS) | 3-4 | NTN integration with optical backhaul and undersea access |
| OWC/LiFi/OCC | kb/s to ~10 Gb/s peak | ms | meters to 10s of meters | IEEE 802.15.7-2018 IEEE 802.11bb | 5-6 | 2-3 | Integration for joint communications, sensing and positioning |
| Maritime (surface) | kb/s-low Mb/s (VHF/satellite/ microwave); Gb/s over short-range FSO | ms (RF/ microwave); 20-100s ms (LEO/GEO satellite) | ~20-50 km (line-of-sight); global (satellite) | IMO, ITU-R (maritime mobile service), IALA | 8-9 | 4.-5 | Reliable handoff between surface links and NTN under variable sea state |
| Undersea (subsurface) | kb/s (acoustic); up to 10 Gb/s (UOWC, <100 m, experimental) | 0.67 s/km (acoustic); µs (UOWC) | 10s km (acoustic); <100 m (UOWC) | NATO STANAG 4748, ISO TC43/SC3, IEEE OES | 7-8 (acoustic); 4-5 (UOWC) | 2-3 | Surface-interface gateway design under real sea-state conditions |

This paper argues for treating the undersea, maritime, terrestrial, aerial and space domains as a single integrated design space and sets out the research agenda needed to realise that vision, a vision that is shared with the UK national hub on network of networks (TITAN). Realising that vision requires cross-fertilisation between research communities that have developed in parallel for long enough that they no longer share vocabulary, channel models, or protocol assumptions, let alone standards. We propose an all-photonic network backbone as the unifying physical infrastructure, with AI-driven orchestration as the cross-medium control layer. We also identify the open challenges across each technical pillar that the field must address to make this feasible. Section 2 provides a structured review of the state-of-the-art across all five domains and the existing cross-domain integration literature, from which the structural gap described above is substantiated in detail. Section 3 develops the convergence framework and its technical pillars. Section 4 sets out a research roadmap from near-term testbed validation through cross-medium field trials to global-scale deployment.

The policy context is favourable for exactly this kind of work. Focusing on the UK, the Digital and Technologies Sector Plan commits £240 million to advanced connectivity technologies, naming non-terrestrial networks and AI integration as explicit research priorities [19]. A UK-Japan Industrial Strategy Partnership has been established within the same framework, creating bilateral conditions for the kind of coordinated international research that cross-medium integration requires [20].

## II. Built in Isolation

The six domains that together could form a heterogeneous multi-medium network have each developed their own mature research community, standards infrastructure, and performance trajectory. What follows is not a comprehensive survey of each field, which would require a standalone paper in its own right, but a characterisation sufficient to establish where each stands, where each is heading and why the trajectory of each does not naturally lead toward cross-medium integration. A comparison across domains is summarised in Table 1.

### *A. Optical fibre and photonics*

Optical fibre networks form the backbone of the communication infrastructure and continue to achieve the highest absolute increases in transmission capacity. Deployed transoceanic systems operate in the multi-Tb/s range per fibre pair, driven by successive generations of coherent detection, advanced modulation formats and digital signal processing (DSP) improvements that have brought the Shannon limit of conventional single-mode fibre within reach [21]. Laboratory demonstrations have pushed further still: a demonstration of 22.9 Pb/s in a single fibre at short distances illustrates the capacity available through space-division multiplexing [22]. Closer to deployment reality, NICT and UCL demonstrated an ultrawideband 450 Tb/s over a legacy 39 km metropolitan link using SMFs [23] covering multiple fibre transmission showing that the headroom remaining in standard commercial fibre plant is still substantial.

The direction of travel within the fibre community is toward energy efficiency as much as raw capacity. The all-photonic network concept, developed through NTT's Innovative Optical and Wireless Networks (IOWN) initiative and supported by NICT's Beyond 5G research programme, proposes eliminating optical-to-electronic conversion at intermediate network nodes, preserving end-to-end single-wavelength paths from source to destination. IOWN 1.0, the all-photonic network (APN), is commercially deployed in Japan. The goals are a 100-fold reduction in power consumption, a 125-fold increase in bandwidth, and a 200-fold reduction in end-to-end latency relative to conventional packet-switched infrastructure [24]. Standardisation is progressing through the IOWN Global Forum, which now counts over 150 member organisations including major hyperscalers and network equipment vendors.

What the fibre research community has lightly addressed is the interface between the photonic domain and the other media that a heterogeneous network would require it to connect to. Every link in a multi-medium path that involves RF, acoustic or FSO segments introduces an optical-to-electronic conversion that the APN concept is designed to eliminate, illustrated in Fig. 3. Designing photonic interfaces that preserve the APN's advantages up to and across those conversions remains an open problem, and one the fibre and photonics community has little institutional incentive to solve on its own. **Open challenge: photonic interface design for multi-domain handoff without sacrificing APN energy and latency gains**.

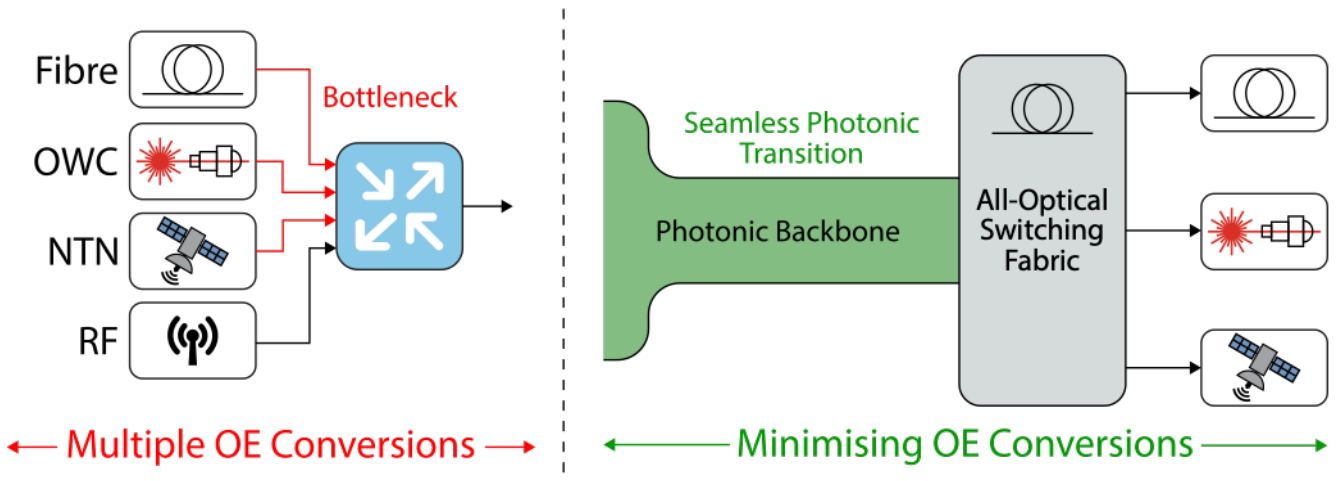

**Fig. 3: The photonic thread: all-optical integration as a backbone to support other mediums. The minimisation of EO and OE conversions is key to reducing power consumption in the network.**

### *B. Free-space optical and terrestrial wireless*

FSO communications has matured considerably over the past decade, driven by two largely separate application areas: (*i*) terrestrial last-mile and building-to-building links, and (*ii*) space-to-ground optical communication for satellite and deep-space systems, which is covered in Section 2.4 alongside the broader NTN landscape.

Commercially available solutions can offer up to 20 Gbps full-duplex connectivity over line-of-sight distances up to 20 km. Coherent WDM techniques borrowed from the fibre domain have transformed what terrestrial FSO can achieve at kilometre scale. A field trial running at a speed of 5.7 Tb/s over a 4.6 km field-deployed link spanning the city of Eindhoven was demonstrated using a 1.1 THz-wide WDM signal and standard coherent transceivers [25]. An earlier field trial achieved 4 Tb/s over 1.8 km with turbulence mitigation and forward error correction (FEC) optimisation [26]. These data rate figures are approaching the capacity of short-haul fibre links, which changes the conversation about where FSO fits in a heterogeneous network from niche last-mile technology to a credible high-capacity wireless segment. The practical constraint is availability. The same 4.6 km Eindhoven link showed 90% availability over a 12-day measurement period including slow fading events, rising to 99% when slow fading was excluded [27]. High data rate and commercial FSO tend to use infrared technology originally developed for fibre optics, taking advantage of high technology maturity. Research on FSO also extends to other wavelengths, *e.g.* near 500 nm to target low-loss windows for UOWC, or below 280 nm for solar-blind operation.

The fundamental challenge for FSO has always been atmospheric impairments: fog, rain, turbulence, beam wander, and beam divergence can reduce link availability dramatically, particularly at longer terrestrial ranges. Hybrid RF/FSO architectures have become the standard proposed mitigation, switching traffic to an RF link when the optical path degrades [14]. This represents the field's most developed cross-technology integration work, but it is an integration of two closely related free-space domains rather than across the broader medium boundary. Most FSO research focuses on optical link performance rather than the interfaces between FSO, fibre and underwater communication systems as part of a full convergent network.

Within the 5G NR standard, sub-6 GHz deployments achieve typical throughputs of 100 Mb/s to 1 Gb/s in real-world conditions, while mmWave deployments reach 4-10 Gb/s per cell under line-of-sight conditions, against a theoretical peak of 20 Gb/s for IMT-2020 compliant systems [28]. The research community is now focused on 6G, where sub-terahertz and terahertz spectrum is under active investigation for access-layer throughputs exceeding tens of Gb/s at short range [29-32].

From a measurement perspective, THz research provides a useful precedent for CHAIN: credible system design at high frequencies depends on scenario-specific propagation measurements and channel models, and existing work has identified measurement methodology and model validation as central research needs [33].

Neither the FSO nor the terrestrial wireless communities have developed channel models or protocol designs that account for the signals arriving from a UOWC link or a satellite NTN segment, and the situation is even worse for maritime links. The assumption embedded in most hybrid FSO/RF work is that both ends of the link are either terrestrial or in free space. **Open challenge: FSO channel modelling and link budget design across the air-water and air-fibre interfaces, where medium transitions introduce impairment statistics that are not fully** **captured by existing free space and underwater channel models.**

### C. *Short-range OWC* and LiFi

The short-range OWC ecosystem includes visible light communications (VLC), LiFi and optical camera communications (OCC). Short-range OWC systems have advanced significantly over the past two decades, driven mainly by the availability of high-speed semiconductor optical sources and the need for extreme network densification. LiFi, which encompasses both visible and infrared links, is best viewed as a networked implementation of OWC, providing bidirectional connectivity and multi-user access. The confinement and small footprint of optical transmissions enable aggressive spatial reuse and increased area capacity beyond what can be achieved through RF; positioning LiFi as a complementary technology to offload traffic from WiFi and cellular networks, while also offering potential as a standalone technology in environments where RF coverage or use is constrained [12].

Early LiFi demonstrations were predominantly based on visible-light light-emitting diodes (LEDs), but this has lately shifted towards narrow-beam IR lasers, including vertical-cavity surface-emitting lasers (VCSELs), which offer tens of GHz of modulation bandwidth, high power conversion efficiency, and narrower beams for higher directionality. This high directionality, however, implies that alignment is an issue, and even modest user mobility, device rotation, or hand/body blockage can interrupt the link entirely. This necessitates dynamic beam-steering and tracking mechanisms capable of following a user or device in real time. Diffractive optics, steerable laser arrays, and reconfigurable photonic beamformers have been proposed as means of maintaining alignment, though each introduces its own trade-off between steering range, switching speed, and system complexity. Experimental demonstrations have achieved up to 42.8 Gb/s per beam over 3 m using wavelength-controlled diffractive beam steering, while an arrayed waveguide grating router (AWGR)-based architecture demonstrated 112 Gb/s per beam with a projected aggregate throughput of 8.9 Tb/s across 80 beams [34, 35]. Silicon-photonic beam steering has also demonstrated 12.5 Gb/s error-free transmission over 1.4 m, highlighting the potential for compact integrated implementations [36]. The complexity, cost, and scalability of precision beam-steering architectures remain significant barriers to large-scale indoor deployment.

Recently, the concept of a programmable wireless environment, powered by reconfigurable intelligent surfaces (RIS) [37, 38] has been proposed as a complementary, largely theoretical approach to the alignment problem: rather than relying solely on transmitter-side beam steering, optical RIS elements embedded in walls or ceilings could in principle passively redirect and reshape light paths to maintain or restore a link when the direct path is blocked or misaligned. This idea shifts part of the alignment burden from the end devices to the environment itself and has been proposed conceptually as a route to combining the high spectral efficiency of narrow-beam LiFi with the robustness expected of a deployable indoor network, though experimental validation of optical RIS at this scale remains limited.

The standardisation position for LiFi has changed materially in recent years with the release of IEEE 802.11bb as the first global standard for light communications, defining LiFi physical and MAC layers designed to operate alongside the established IEEE 802.11 Wi-Fi family, allowing LiFi access points to integrate into existing enterprise wireless infrastructure rather than requiring a parallel network. This marks a significant milestone for the LiFi market, moving the technology from proprietary point solutions toward an interoperable, vendor-neutral deployment path. As it stands, the standard defines a viable and interoperable baseline for commercial deployment, but there is currently no defined path for the advanced DSP techniques driving LiFi's performance ceiling to be incorporated into a future revision of the standard. Looking at the LiFi market, commercial products have progressed considerably. The first commercial LiFi luminaire brought to market by pureLiFi in 2016 offered 42 Mb/s downlink and uplink, serving up to eight simultaneous users. A decade on, pureLiFi's current High-Bandwidth Architecture, unveiled at MWC 2026, delivers low-latency LiFi connections at 10 Gb/s per link [39].

The OCC is employing image sensors rather than photodiodes as receivers [40] and, as such, it provides a unique combination of communication, sensing capabilities and localisation [41] which may prove particularly valuable at cross-domain boundaries, including vehicle-to-vehicle (V2V) communication, maritime navigation, UAV coordination and underwater-to-surface gateway systems [42]. Lately, implementation of OCC has been improved significantly with the use of deep learning [43] while high-speed cameras with micro-LEDs have achieved data rates in Mb/s range [44]. While the technology is maturing mostly within indoor applications, its role within heterogeneous multi-domain networks remains largely unexplored and represents an important direction for future CHAIN research.

**Open challenge: As with the other domains surveyed here, short-range OWC/LiFi/OCC research has developed with minimal engagement with the adjacent photonic domains it should, in principle, integrate with most naturally.**

### *D. RF communications*

RF deserves separate treatment because it is the only communication medium considered in this paper that spans all five domains, undersea, terrestrial, maritime, aerial and space, rather than being largely confined to a single medium. Terrestrial cellular and Wi-Fi are both RF, while current NTN satellite downlinks predominantly rely on RF links, operating in L, S, Ka and Ku bands. Underwater RF communication is limited to extremely low frequencies, where long wavelengths capable of penetrating seawater require very large antennas and support data rates limited to tens of bits per second, used almost exclusively for one-way communication with submarines [10, 45]. RF also provides the bridge in most practical underwater-to-surface relay systems, where an acoustic modem surfaces a signal that then continues as RF.

This cross-domain presence makes RF the most natural candidate for the integration layer in a heterogeneous multi-medium network at the edge of a fibre network. The difficulty is that RF's characteristics vary so substantially across these deployments that they share little beyond the fundamental modulation formats. The sub-millisecond latency of a 5G mmWave link and the multi-second propagation delays of a deep-sea extremely-low-frequency transmission are the same physical phenomenon at opposite ends of a wide parameter space, making unified protocol design challenging. The various forms of RF communications benefit from the most mature standards ecosystem among the domains considered here, spanning 3GPP (5G NR and NTN), ITU-R spectrum coordination across all frequency bands and IEEE 802.11 for local area applications. That maturity is itself part of the isolation problem. The standards infrastructure is deeply optimised for RF-within-RF interoperability and has little mechanism for defining interfaces with domains that operate on fundamentally different physical principles. An additional opportunity at the RF boundary is to make the propagation environment itself reconfigurable. RIS and electronically controlled beamforming front ends can steer or reshape RF propagation without introducing a conventional active relay; real-world testbeds have demonstrated programmable sub-6 GHz reflection and calibrated large array mmWave beamforming [46, 47]. Within CHAIN, these technologies are best viewed as programmable interface elements at terrestrial, maritime and NTN joints rather than as a separate domain. A facility-level research question is how such controllable RF interfaces can be jointly measured and orchestrated with FSO, fibre and underwater links, including whether they can improve resilience during cross-medium handoff and create repeatable boundary conditions for channel characterisation.

**Open challenge: RF protocol and interface design that spans from IoT-class low-rate links through to broadband NTN backhaul within a common cross-medium addressing and orchestration framework.**

### *E. Non-terrestrial networks*

NTN research has seen rapid development over the past five years, driven by the deployment of LEO mega-constellations and the standardisation effort in 3GPP. Commercial LEO constellations now provide connectivity up to 350 Mb/s on business plans with latencies between 25 and 70 ms [48, 49]. Upcoming satellite generations target throughputs beyond 1 Tb/s [50]. Optical inter-satellite links are now a core part of LEO backhaul architecture driven by size, weight and power (SWaP) advantages that are fundamentally enabled by the smaller wavelength and therefore improved diffraction limit compared to RF. For example, Starlink operate over 9,000 space lasers moving approximately 42 petabytes of data per day by early 2024 [51], a development that makes the NTN domain increasingly photonics-adjacent and relevant to the all-optical backbone argument developed in Section 3.

At the other end of the NTN spectrum, HAPS operating at 20-25 km altitude offer propagation latencies below 1 ms with ground coverage footprints of up to several hundred kilometres radius, though commercial deployment remains at early stages [52, 53]. HAPS deployment can also be rapid, and the onboard system can be regularly upgraded, contrary to satellites.

3GPP Release 17 and 18 established the normative framework for satellite and HAPS integration into 5G NR, as described in Section 1. The Release 19 and 20 work programmes are extending this to GNSS-resilient operation, enhanced inter-satellite handover, and IoT-NTN Phase 4 capabilities [54] deepening standardisation within the terrestrial-to-space segment without expanding its scope to adjacent domains.

The scope boundary is the critical limitation. Integration with undersea systems, all-optical backhaul, and the acoustic domain does not appear in the 3GPP work programme. The implicit assumption throughout is that the ground segment connects to a conventional terrestrial or maritime network.

**Open challenge: NTN integration with all-optical backhaul and with undersea access domains, including handoff protocol design across the satellite-surface-subsea boundary.**

### *F. Underwater and Maritime communications*

Maritime communications address the challenge of providing reliable connectivity across the 71% of the Earth's surface covered by oceans, connecting vessels and platforms, the remoteness of which often makes wireless communication the only feasible solution for connectivity.

Point-to-point, above-sea maritime OWC has much in common with point-to-point FSO domains, albeit with a general intensification of challenges arising from atmospheric effects (fog, humidity, turbulence) as well as pointing-and-tracking, owing to sea-wave-induced stochastic vessel motion. These challenges make hybrid RF/FSO solutions particularly attractive for maintaining uptime during unfavourable conditions. Whilst blue water maritime communication makes heavy use of NTN connectivity, ship-to-ship and ship-to-shore communication is particularly relevant to ships at port or in formation. In these contexts, FSO can extend the data rates enjoyed onshore to vessels at port, further the automation of surface and sub-sea vessels and even play an important role in positioning. The combination of a similar but distinct challenge space to other point-to-point links as well as the need for both TN and NTN connectivity is a clear indicator of the heterogeneous nature of maritime connectivity and one that is strengthened further when considering the frequent role of surface vessels as relays for underwater communication to sub-sea vessels themselves.

Maritime communication is particularly notable for its complete lack of specific FSO standards, despite a long history of marine standardisation and the existence of relevant standards such as [55] for seakeeping and [56] for navigation and RF communication.

Underwater communication is the least visible of the five domains in the literature and carries the largest gap between its current capabilities and what a heterogeneous multi-medium network would require.

Acoustic communication is the historically dominant deployed technology for range beyond a few tens of metres. Acoustic waves propagate at approximately 1,500 m/s in seawater, producing one-way propagation delays of around 0.67 s/km, and achieve data rates from a few kb/s at long ranges to tens of kb/s at moderate ranges [57]. The bandwidth-distance product of underwater acoustic channels falls several orders of magnitude below any of the other media in this paper.

UOWC addresses the bandwidth limitation but at the cost of range. Experimentally, data rates of 10 Gb/s have been demonstrated at distances of 28-38 m in coastal water using laser diodes [58]. Real-time LED-based systems have reached 135 Mb/s in pool experiments [59]. These figures are approaching the data rates of short-range terrestrial wireless systems at a fraction of the range.

Simulation studies project that spatial diversity-assisted UOWC could extend Gb/s-class links to ranges exceeding 1,000 m in clear ocean water [60]. However, a substantial gap remains between laboratory demonstrations and real-world deployment, where water turbidity, absorption, scattering, random temperature variations, and environmental variability degrade link performance. The practical operating regime for UOWC today is therefore secure, short-range, high-rate communication between platforms in proximity, with acoustic modems handling longer-range, lower-rate links. Bridging those two regimes within a single coherent system and then connecting that system to a surface gateway and onward to NTN or terrestrial infrastructure remains an open engineering problem with no deployed solution.

Standards for underwater communications are fragmented across several bodies without the coordinating authority of 3GPP. NATO's JANUS standard, STANAG 4748, provides the first international standard for underwater acoustic signalling as a common wake-up and coordination protocol, but does not define a full network stack [61]. ISO TC43/SC3 covers underwater acoustics. Institute of Electrical and Electronics Engineers (IEEE) Oceanic Engineering Society (OES) maintains a separate standards activity. None of these bodies has mechanisms for defining interoperability with the RF, FSO or fibre domains that an integrated multi-medium network would require at the surface interface.

The cross-medium challenge is toughest here because the undersea domain is separated from the others not just by protocol incompatibility but by a physical boundary. The air-water interface attenuates RF signals to a few centimetres of penetration at communications frequencies, making direct RF-to-underwater wireless links impossible at practical data rates. It is also very challenging to maintain reliable OWC links through the water surface due to refraction and reflection at the surface, stochastically perturbed by waves. Every realistic multi-medium path that includes an underwater segment must pass through an acoustic-to-optical or acoustic-to-RF conversion at or near the surface, and the design of those conversions under real sea-state conditions, with the latency and reliability implications for end-to-end path performance, has received almost no systematic study. **Open challenge: surface-interface gateway design spanning acoustic, optical and RF domains under realistic sea-state conditions, with end-to-end latency and reliability models for paths that include underwater segments.**

### *G. Cross-Domain Integration and its Limitations*

Cross-domain integration has attracted growing research attention, concentrated mainly on two-domain pairings. Satellite-terrestrial network integration has been studied extensively within the 5G and early 6G communities, addressing handover protocols, interference management and hybrid RF architectures [62]. Relay-assisted hybrid FSO-RF systems have been proposed to extend optical wireless reach through RF backup under atmospheric disruptions [14], and UAV-assisted relay architectures have been investigated for coverage extension and emergency communications [15]. In each case, the integration work takes two domains and holds the rest of the network as either terrestrial or fixed. There have also been numerous studies of RF transmission over FSO [63-67].

The most developed attempt at treating more than two domains simultaneously is space-air-ground integrated networks (SAGIN), which proposes satellite, aerial and terrestrial segments as a unified architecture. The foundational SAGIN survey [68] was the first to address all three segments together rather than space-ground or air-ground pairings in isolation. Several follow-on surveys and architectural proposals have extended the framework into 6G contexts with machine learning-based resource allocation [69, 70].

The research effort behind SAGIN is substantially lower than the literature behind any of its constituent two-domain pairings. The satellite-terrestrial body of work alone is larger in volume than the entire SAGIN literature, and the work within SAGIN is dominated by analytical and simulation studies: coverage probability derivations using stochastic geometry, resource allocation algorithms evaluated through Monte Carlo simulation, and architectural proposals validated against simulated scenarios. No over-the-air field trial spanning all three SAGIN domains simultaneously has been reported in the open literature to the best of the authors' knowledge. SAGIN has produced useful theoretical foundations, but the experimental and standards work has not followed.

The most recent extension, space-air-ground-sea integrated networks (SAGSIN), incorporates surface maritime stations, tethered balloons and HAPS relays to derive coverage probability for users on the open ocean surface [71]. The results show that adding maritime relays improves coverage for users far from the coastline, which terrestrial and satellite systems alone cannot serve reliably. The subsurface domain does not appear in the framework. The architecture is RF-centric throughout, with no photonic backbone and no AI orchestration layer, and the most recent SAGSIN work acknowledges that sea-surface optical and underwater acoustic links face bandwidth and power constraints that the existing framework does not resolve [72].

SAGIN and SAGSIN show that the field is moving towards multi-domain integration and has started to build the analytical tools for it. The problem is that the trajectory has not yet produced a comprehensive experimental platform, a multi-domain communication system, or a cross-domain standards effort. The CHAIN framework proposed in Section 3 extends it to five domains, adds the all-photonic network backbone as the unifying physical infrastructure, and introduces AI-driven orchestration as the cross-domain control layer. Achieving CHAIN would enable real-world and repeatable testbed interface testing for the first time, globally. Developing this before SAGIN has consolidated matters because the architectural decisions made early in a framework tend to persist through later generations of work, making them progressively harder to revise.

### *H. Cross-Domain Gaps*

The pattern across sections 2A-2G above is consistent such that each domain's internal research is mature, two-domain pairings have produced substantial analytical and some experimental results, and SAGIN and SAGSIN have extended the ambition to three and four domains without yet closing the gap to deployed systems. What remains unaddressed in all this work is a unified treatment of all domains simultaneously, including the subsurface, with a photonic backbone and a cross-domain orchestration layer.

Closing that gap starts with channel characterisation. Signal impairments in multi-domain paths do not compose additively from the impairments of each segment in isolation. Multi-domain propagation introduces effects that are not captured by existing channel models and require measurements. How statistics evolve across the air-water interface, the optical-to-RF conversion boundary, and the fibre-to-FSO junction within a single end-to-end path has not been addressed in the literature. These models cannot be derived by composing within-domain models, and they are a prerequisite for any principled CHAIN system design.

Above the physical layer, transport protocols become the limiting factor. A path spanning fibre, 5G wireless, a LEO satellite hop, and an underwater acoustic segment encounters several orders of magnitude of latency variation within a single connection. Protocols designed for terrestrial networks handle perhaps one order of magnitude. Adapting or replacing them for this range is a research problem the SAGIN and two-domain literatures have not addressed, because each implicitly assumes the rest of the path is handled elsewhere.

Orchestration is the hardest gap to close precisely because it depends on the other two. Real-time routing across various domains with different availability, capacity and failure characteristics requires a control plane with physical-layer awareness of conditions in all five simultaneously, including underwater optical clarity and sea state. SAGIN and SAGSIN have proposed architectural frameworks for multi-domain orchestration, but without channel models to inform routing decisions or protocols to implement them, those frameworks have no operational basis.

Standards fragmentation compounds all three. 3GPP, ITU-R, ITU-T, NATO STANAG 4748, ISO TC43/SC3 and IEEE OES each govern their own domain with no coordination mechanisms for the boundaries among them. Until a shared reference architecture provides common abstractions above those domain-specific standards, deployments that cross domain boundaries will continue to require bespoke integration at every joint. The CHAIN framework introduced in Section 3 proposes that architecture.

## III. Designing for Convergence

### *A. The Unified Design Space: The CHAIN Framework*

The survey in Section 2 points to a consistent pattern. Each domain has optimised within its own boundaries, two-domain integration has produced useful but bounded results, and SAGIN has begun to address three domains analytically without reaching the experimental or standards work needed to realise the concept. The most recent SAGSIN extension has added the ocean surface without addressing the subsurface, and neither includes a photonic backbone or an AI orchestration layer as architectural elements.

This paper proposes treating the undersea, maritime, terrestrial, aerial and space domains as a single integrated design space. We coin this the CHAIN framework: Connected Heterogeneous All-medium Integrated Network. The framework treats the five domains as a single design space from the outset, with the joints between domains as first-class engineering problems rather than integration afterthoughts. Three questions follow from this framing. What provides the physical infrastructure layer that makes the five domains interoperable? How does the network make real-time decisions about routing, handoff and resource allocation across domains with radically different characteristics? And what concrete advantages does this design space offer over five independently optimised systems? The following subsections begin to address each in turn.

### *B. The All-Photonic Network Backbone*

The physical infrastructure candidate that best fits the CHAIN framework is the APN, developed through NTT's IOWN initiative. The APN replaces conventional optical-electronic-optical conversion at intermediate network nodes with end-to-end photonic processing, preserving signals in the optical domain from source to destination wherever the path permits. The APN IOWN 1.0 service launched commercially in Japan in March 2023 through NTT, targeting up to 200 times lower end-to-end latency relative to conventional packet-switched networks in commercial operation [24]. The broader IOWN roadmap targets 100x improvement in power efficiency, 125x increase in transmission capacity, and the 200x latency reduction across successive generations through to 2030.

For the CHAIN framework, the APN's relevance extends beyond its capacity and efficiency targets. Optical fibre, FSO terrestrial links and inter-satellite optical links already form a physically continuous optical path from the terrestrial network through the atmosphere to low earth orbit. In principle, the APN allows that path to remain in the optical domain throughout, with IOWN Global Forum standardisation supporting interoperability across the segments. NTT demonstrated a 2,893 km APN connection between Tokyo and Taiwan via Chunghwa Telecom in August 2024, showing the architecture is viable at transoceanic distances [73]. Looking further ahead, NTT's broader 6G-IOWN 'Extreme Coverage Extension' infrastructure initiative explicitly integrates large-capacity underwater acoustic communications. Running in tandem with the IOWN 3.0 timeline targeting a 125x capacity scaling by 2029, this expansion into the maritime domain signals that the photonic backbone concept is already evolving in the direction the CHAIN framework requires.

The limit of the APN is where CHAIN's integration challenge becomes most concrete. UOWC links operate at comparable wavelengths to terrestrial FSO systems, but the air-water interface introduces absorption and scattering that prevent a direct photonic path from continuing below the surface. RF links, which dominate NTN access and much of the terrestrial wireless layer, require optical-to-electronic conversion at every interface with the photonic backbone. The APN therefore provides the unifying infrastructure for the fibre, FSO and optical satellite segments of CHAIN, while the undersea and RF segments require interface designs that preserve as much of the APN's efficiency advantage as possible across the conversion boundary.

Photonic generation and fibre distribution of mmWave signals provide one established route across the optical-RF boundary. Prior radio-over-fibre demonstrations have shown that high-frequency RF waveforms can be generated in the optical domain, distributed over fibre and recovered for wireless transmission [74]. The CHAIN-specific challenge is to quantify the energy, latency, dynamic-range and signal-fidelity penalties of such interfaces when they are embedded within a multi-medium path rather than treated as a standalone access link.

### *C. AI-Driven Orchestration*

The APN backbone addresses the physical infrastructure question. How the network makes decisions across five domains with radically different latency, availability and capacity profiles requires a control layer that operates above the domain-specific standards. Within CHAIN, this function is assigned to an AI-driven orchestration layer implementing intent-based networking principles: the network is given operational objectives such as minimum end-to-end latency, maximum reliability, or energy efficiency, and determines the routing and resource allocation required to meet them across whichever combination of domains is available at a given moment [75].

Digital twin representations of the cross-domain network state provide the real-time sensing input the orchestration layer requires. Each domain's physical conditions, underwater optical clarity, atmospheric turbulence on FSO links, satellite orbital geometry and available bandwidth, terrestrial network load, etc., are continuously modelled and fed into routing decisions. Digital twins are established in terrestrial network management and are under active development for NTN systems [76, 77]. The open problem is a digital twin architecture that spans all five domains simultaneously, with the latency and synchronisation requirements that cross-domain real-time routing imposes.

A further constraint specific to CHAIN is that underwater nodes operate under severe bandwidth limitations that preclude participation in centralised AI training. Federated learning, already proposed for SAGSIN applications to address this constraint [78], distributes model training to the network edge, with local models trained on available data at each domain and aggregated at the cross-domain orchestrator without requiring raw data transfer from bandwidth-constrained subsurface nodes. How federated learning performs under the intermittent connectivity and extreme latency asymmetry of the acoustic domain is an open research question.

The orchestration layer remains deliberately thin in this framework. Its function is coordination rather than computation, selecting paths and allocating resources across the domain-specific management systems that each continue to handle their own physical layer optimisation. This preserves the engineering maturity of each domain's existing standards infrastructure while providing the cross-domain decision layer that currently does not exist.

### *D. The Convergence Advantage*

The case for treating five domains as one design space rests on five claims that integration (a) improves resilience, (b) extends coverage, (c) increases efficiency and (d) enables performance optimisation across workloads. The two-domain integration literature provides partial support for each. For the five-domain case, these remain hypotheses, substantiated by physical reasoning and analogy to two- and three-domain results, and the research agenda in Section 4 is partly defined by the need to test them.

***Resilience:*** Hybrid FSO-RF systems provide the clearest two-domain precedent. When atmospheric conditions degrade the optical link, traffic can be switched, an RF link, improving overall link availability. Because the two technologies are affected differently by atmospheric conditions, combining them significantly improves link availability achieving 99.999% compared with isolated technology approaches where neither link alone could guarantee service [79]. Mentioned above, the availability study of the field-deployed 4.6 km Eindhoven FSO link, operating alongside an RF fallback, reported 90% availability over 12 days including slow fading events, rising to 99% when slow fading was excluded [27].

In the satellite-terrestrial domain, integrated NTN architectures have the potential to extend coverage and serve unconnected users when connectivity is unavailable and provide an alternative communication path during infrastructure failures and disaster scenarios that route traffic through LEO satellites when terrestrial base stations are unavailable maintain service continuity under outage scenarios that would otherwise leave users unserved [2]. A CHAIN network with undersea acoustic, terrestrial, aerial and space links all available to the orchestration layer has more path diversity than any two-domain combination, and a higher probability of maintaining connectivity under any single-domain failure. Quantifying the gain from technology combination requires the cross-domain path models identified in Section 2.7 and cannot yet be made precisely.

***Coverage:*** The SAGIN literature has established analytically that combining satellite, aerial and terrestrial segments extend coverage to areas no single segment can serve independently [68, 69]. The SAGSIN extension demonstrates that incorporating maritime surface relays into the SAGIN architecture provides non-trivial coverage gains for users far from the coastline, well beyond the footprint of either terrestrial networks or standard satellite ground segments [71]. The CHAIN framework extends this to the subsurface domain: underwater optical and acoustic links reach environments inaccessible to any of the SAGSIN domains, and their integration is the necessary step for coverage that includes the ocean volume, not only its surface.

***Efficiency:*** As mentioned, the IOWN roadmap targets a 100x improvement in power efficiency relative to conventional infrastructure through end-to-end photonic processing across the IOWN programme through to 2030 [80]. IOWN 1.0's commercial deployment has already demonstrated the 1/200 latency reduction target in live operation, confirming the architecture's core energy-efficiency argument at the first generation. In the FSO-RF hybrid literature, the availability of a high-capacity optical primary link means the lower-capacity RF fallback carries traffic only under degraded conditions, reducing the average energy cost per bit relative to an RF-only system [14]. For a CHAIN network, shared orchestration infrastructure and common interface standards would avoid the duplicated engineering effort that currently accompanies every multi-domain deployment, though the quantitative gain from this coordination has not yet been modelled.

***Performance optimisation:*** A three-medium hybrid FSO/RF/THz relay system demonstrates analytically that an adaptive combining scheme across three media achieves better outage probability and throughput than any single medium can provide [81]. The performance gain scales with the diversity of available media because different media have complementary impairment statistics: atmospheric absorption degrades FSO links but not acoustic links; Doppler spread degrades acoustic links but not fibre. In a CHAIN network, the orchestration layer can assign workloads to the best-fit medium, low-latency traffic to the lowest-latency available path, high-reliability paths for mission-critical sensor data, high-throughput satellite links for bulk transfer from offshore platforms. This workload-to-medium matching is possible only when all five domains are available to a common scheduler, and its performance advantage over domain-specific scheduling has not been demonstrated at five-domain scale.

### *E. Cross-Medium Physical Layer*

Realising the CHAIN framework depends on channel models and link budget tools that do not yet exist for cross-domain paths. Within individual domains, channel characterisation is well established with models describing propagation effects for fibre, FSO (for horizontal and vertical links), terrestrial and maritime RF, opto/acoustic underwater communications.

Within individual domains, channel characterisation is well established, with models describing the dominant propagation impairments for fibre, FSO, RF and underwater communications [9, 10, 71, 82, 83]. The problem is that these models are defined within domains and do not describe how signal statistics evolve across the transitions between them.

At the air-water interface, a signal arriving from an FSO or RF link encounters a medium boundary where refractive index, absorption, and scattering properties change discontinuously. The compound statistics of a path that includes an underwater optical segment, a surface relay, and an FSO segment to a HAPS or satellite have no published characterisation. Similarly, the end-to-end performance of a path spanning fibre, terrestrial FSO, and an inter-satellite optical link cannot be predicted by composing the individual domain models, because the correlation structure of impairments across the domain boundaries is unknown. This is not merely a modelling inconvenience. Link budget design for a CHAIN network path requires end-to-end noise and fading characterisation to set power levels, modulation orders, and error correction parameters. Without cross-domain channel models, those design decisions cannot be made on a principled basis, and the CHAIN framework remains conceptually complete but practically unrealisable. Developing these models, through measurement and channel sounding campaigns at domain interfaces under realistic conditions, is the physical layer priority for the research agenda. Equally important is the creation of open, well-curated reference datasets that capture cross-domain propagation characteristics. These datasets would provide benchmark data for model validation, support reproducible research, enable the development of AI-driven orchestration algorithms, and serve as a shared resource for both academia and industry that aligns with the upcoming OWC facility objectives.

For the RF portion of this measurement programme, the facility should support calibrated sub-6 GHz and mmWave channel sounding with configurable beamforming and programmable-surface states [46, 47], while retaining a pathway to sub-THz/THz measurements as instrumentation matures [33]. Cross-domain datasets should therefore record not only environmental conditions but also beam state, RIS configuration and calibration metadata, allowing physical-layer measurements to be reproduced and consumed by the digital-twin and orchestration layers.

### *F. Standards and Interoperability*

The CHAIN framework requires a set of common abstractions above the domain-specific standards that currently govern each medium separately, and the standards gap has been discussed above. Building a shared reference architecture above the five domain-specific standards is a harder problem than previous interoperability efforts because the domains differ from each other not just in protocol but in fundamental physical assumptions about node mobility, round-trip time, and link reliability. A handoff protocol that works between terrestrial cellular and LEO satellite assumes both endpoints can exchange a sequence of messages within a second or two. Adding an underwater acoustic segment to that handoff puts one endpoint several seconds away in propagation time alone, which breaks the protocol at a level that cannot be patched by parameter adjustment. The scope of the IOWN Global Forum standardisation effort, even with formal ITU collaboration, stops well short of the acoustic and RF domains that CHAIN requires. Until a body exists that can define interface primitives across all five simultaneously, CHAIN deployments will require bespoke integration at every domain boundary.

Extending the IOWN model to include the underwater domain and the RF access layer and providing the addressing and handoff primitives that the CHAIN orchestration layer needs to function, is the standards work required to complement the physical layer and protocol research.

Security across heterogeneous trust domains compounds the standards challenge. Each of the five domains operates under different threat models: physical cable tapping for fibre, atmospheric eavesdropping for FSO, jamming and spoofing for RF and satellite, maritime eavesdropping, and acoustic side-channel attacks for underwater systems. A CHAIN network must maintain end-to-end security assurance across trust domain boundaries that its constituent standards were not designed to span. The GCOT 6G security and resilience principles provide a starting point for the space and terrestrial segments [84], but their extension to the undersea domain requires new work

### *G. Core Open Problems*

The preceding sections point to a set of problems that sit across domain boundaries. The most fundamental is channel characterisation across domain transitions. Within each domain, channel models are mature and experimentally validated: Gamma-Gamma and generalised Gamma distributions describe optical turbulence included scintillation above and below the surface, Thorp's formula and its successors describe acoustic attenuation as a function of frequency and range, and well-established path loss models cover the RF segments. What does not exist is a treatment of how signal statistics evolve across the transitions between these domains within a single end-to-end path. The air-water interface is where this gap is sharpest. An optical signal crossing from air to water encounters wave-induced beam wander whose statistics depend on sea state, described by the Joint North Sea Wave Project (JONSWAP) spectrum [85], and whose coupling into the received intensity distribution is not captured by either the atmospheric turbulence models above or the absorption-scattering-turbulence product models below. A recently proposed Monte Carlo ray-tracing approach models the water-to-air channel incorporating both sea surface statistics and Unmanned Aerial Vehicle (UAV) platform instability [86], but this remains a two-segment treatment. A compound model that propagates signal statistics from the underwater source through the surface interface, across the atmospheric segment to a HAPS or satellite, and then through inter-satellite optical links to a fibre endpoint, accounting for the correlations between impairments at each boundary, does not exist and cannot be constructed by composing the within-domain models.

Without those models, the protocol problem cannot be properly framed. Transport protocols assume a statistical model of path behaviour, even when that model is implicit. Transmission Control Protocol (TCP)'s congestion control infers available bandwidth from packet loss and round-trip time, and its congestion window grows and collapses on timescales calibrated for round-trip times (RTTs) in the range of tens to hundreds of milliseconds. The Quick User Datagram Protocol (UDP) Internet Connection (QUIC) protocol's bottleneck bandwidth and round-trip propagation time and CUBIC algorithms extend the viable Use Real-Time Text (RTT) range but remain calibrated for paths where the round-trip time is stable within an order of magnitude [87]. A CHAIN path that includes a fibre segment (RTT in the microsecond range per kilometre), a 5G hop (single-digit milliseconds), a LEO satellite link to a vessel at sea (around 50-80 ms), and an underwater acoustic segment at 10-50 km range (one-way propagation of 6-33 seconds, giving RTTs of 12-67 seconds) spans seven orders of magnitude of propagation delay within a single connection. At acoustic timescales, TCP's retransmission timeout fires repeatedly before the first acknowledgement arrives, collapsing the congestion window to one segment and effectively halting transmission. QUIC's acknowledgement frequency mechanism reduces this problem for satellite links [88] but has not been evaluated at acoustic RTTs. The design of a transport layer that treats this range as a parameter rather than a pathological edge case is an open problem, and it depends on the cross-domain channel statistics that the measurement programme above would produce.

The photonic interface problem is independent of the channel modelling work but equally foundational. The APN backbone's latency and energy advantages derive from keeping signals in the optical domain and avoiding opto-electrical/electro-optical (OE/EO) conversions at intermediate nodes. The wavelength used in fibre transmission and inter-satellite optical links is 1550 nm, in the near-infrared C-band. Water absorbs this wavelength almost completely within a few centimetres, so the CHAIN path cannot continue optically below the surface without a wavelength conversion into the blue-green transmission window of 400-530 nm where seawater is most transparent. This wavelength conversion from 1550 nm to the blue-green range and back introduces additional OE and EO stages that the APN architecture does not currently address. The power and latency penalty of that stage occurs at a submerged or surface-mounted gateway under the mechanical constraints of a marine environment and platform motion which has not been fully characterised and can be location dependent The convergence approaches under development within the IOWN programme target data centre interconnects and assume stable, controlled environments; their adaptation to outdoor and subsurface deployment is not part of the current roadmap [89].

The orchestration problem, as discussed in Section 3.3, cannot be properly addressed until the channel models exist to populate the digital twins with state information, nor until the protocol designs exist to implement the routing decisions the orchestration layer produces. The additional complication specific to the cross-domain case is that the orchestration layer's own control signalling must traverse the same heterogeneous path it is trying to manage. A routing decision transmitted from the orchestrator to an underwater acoustic node takes the same 6-33 seconds to arrive as the data it is rerouting. Hierarchical control architectures reduce this problem by delegating intra-domain decisions to local controllers, but the timescale mismatch between the fastest-changing domain states (atmospheric turbulence on FSO links, which has a coherence time in the region of a millisecond) and the slowest-reachable nodes (deep acoustic) means the orchestration system must maintain coherent state across processes operating on timescales six orders of magnitude apart. No existing multi-domain orchestration framework can handle this range.

The standards gap is last in the dependency chain because it cannot be resolved until the technical problems above have produced artefacts that are worth standardising. A handoff signalling protocol between the terrestrial cellular and acoustic domains, for example, requires a shared understanding of what constitutes a handoff condition in a domain where round-trip signalling takes tens of seconds and node position changes slowly relative to all other domains in the path. The existing handoff protocols in 3GPP NTN work assume that the relevant state can be exchanged between the handoff participants within a few seconds at most. Extending this to the acoustic domain is not a matter of adjusting timer values; it requires rethinking the handoff model from the ground up for a domain with fundamentally different dynamics. The IOWN Global Forum's ITU collaboration provides a starting point for aligning the optical segment standards, but the underwater domain falls outside both bodies' current scope, and no mechanism exists for the cross-domain interface primitives that CHAIN requires. These are the dependencies the research agenda in Section 4 follows.

## IV. From Testbed-to-Trial Roadmap

The research agenda that follows has a natural ordering. The channel characterisation work must come before the protocol work, which must be performed before serious orchestration design. All three must be complete before multi-domain integration trials can produce meaningful results for full network operation. Therefore, we split the roadmap into near, middle and long-term viewpoints.

### A. Near-Term: Measuring the Boundaries

The immediate priority is the system measurement. Most of the open problems discussed above remain open not because they are technically intractable but because the experimental infrastructure to address them does not exist in the right form. Each of the five constituent domains has its own testbed infrastructure: fibre transmission test networks, acoustic-optical test ranges, maritime platforms, satellite channel emulators and ground stations, FSO rigs. What does not exist is a facility where all five are co-operated as a single system, making the cross-domain boundary conditions, the statistics of the air-water optical interface under realistic sea conditions, the compound impairment of a path that transitions from acoustic to optical to atmospheric to fibre, observable in a controlled and repeatable way. Until such a facility exists, the foundational cross-domain channel models identified above will remain unexplored and unpublished, because the measurements needed to validate them cannot be made. Establishing that co-located multi-domain measurement capability is the prerequisite for everything else in this roadmap.

The specific measurement targets in this stage are the air-water optical interface under varying sea state conditions, characterised against the JONSWAP spectrum and validated against both clear and coastal water profiles; the compound impairment statistics of two-domain paths, initially acoustic-to-optical and FSO-to-fibre, measured jointly rather than composed from separate campaigns; and the photonic interface at the wavelength conversion boundary between 1300 nm/1550 nm telecom-band signals and the 400-530 nm blue-green window required for underwater propagation. These measurements feed directly into the channel models that protocol and interface design require.

A calibrated RF interface suite, initially spanning sub-6 GHz and mmWave and expandable toward sub-THz/THz, should run alongside these optical measurements so that controlled RF/optical handoff experiments and joint-state datasets can be produced [33, 46, 47].

In parallel with the measurement campaigns, protocol design work can proceed in simulation using parameterised channel models drawn from the available single-domain literature as placeholders, with the explicit goal of producing designs that are ready for validation against real measurements as soon as those become available. The transport-layer problem is well enough scoped, extending QUIC's acknowledgement frequency adaptation to acoustic RTT ranges and testing congestion control behaviour under the compound latency variance of two-domain paths, that simulation studies can begin without waiting for the full cross-domain channel characterisation to be complete. Similarly, digital twin architectures for the cross-domain orchestration layer can be developed and tested against synthetic network states before real cross-domain measurement data is available to populate them.

### B. Medium-Term: Building the Joints

As channel models are validated and protocol candidates identified, the medium-term work shifts towards integration. The starting point is two-domain testbed demonstrations, each targeting one of the boundary joints. The goal is not to demonstrate each link in isolation, which has largely been done in the two-domain literature, but to demonstrate that the joints between domains behave as the cross-domain channel models predict. Discrepancies between predicted and observed joint behaviour are the research output at this stage, as much as the demonstrations themselves. This allows closed-loop feedback and co-development across technology interfaces, whilst still performing requirements gathering and distillation.

The next step is three-domain integration. The offshore wind operational context that opened this paper is one example of where such three-domain paths occur in practice, though it is not the only one; ocean climate monitoring, maritime search coordination and submarine cable maintenance present equivalent multi-domain connectivity demands. A three-domain trial conducted in any of these contexts would validate the integration architecture against real operational requirements without requiring the full five-domain system to be in place.

The orchestration layer's digital twin can be validated at this stage against real cross-domain state measurements for the first time. The timescale mismatch between fast-varying FSO states and slow-varying underwater states as an open orchestration problem becomes empirically observable once a three-domain testbed is operating, and the hierarchical control architecture can be evaluated against it.

Standards engagement should begin in this stage rather than waiting for the full five-domain demonstration. The IOWN Global Forum is the natural venue for photonic interface standardisation across the optical segments. Engagement with 3GPP on handoff protocol extensions for high-latency undersea access segments, and with IEEE OES on surface-interface gateway specifications, is most productive when the testbed work provides concrete technical proposals to bring to the standards process rather than conceptual arguments. Link security needs to be considered in this process, including physical layer security and vulnerabilities, their impact on the integrated system, and mitigation strategies.

### C. Long-Term: Testing the Framework

The long-term goal is a full five-domain CHAIN field trial: a deployed system in which undersea acoustic and optical links, maritime, terrestrial and FSO wireless segments, NTN satellite and aerial platforms with an all-photonic fibre backbone operate as a single network under the CHAIN orchestration layer, with real workloads and real cross-domain handoffs. The system does not need to be permanent or large-scale at this stage; a trial deployment of modest geographic scope that exercises all five domain boundaries simultaneously under realistic operational conditions is sufficient to validate the CHAIN architecture as a unified network.

Such a trial would produce the first empirical test of the convergence advantages, whether path diversity across various domains provides resilience gains beyond two-domain switching, whether the APN backbone's efficiency advantage survives the wavelength conversion stages at the underwater and RF boundaries, and whether the orchestration layer's routing decisions improve end-to-end performance relative to domain-specific management. These are the questions that the analytical and simulation literature, including SAGIN and SAGSIN, has posed but not yet answered with field data.

The international dimension of this work becomes most concrete at the field trial stage. A five-domain deployment of sufficient geographic scope to be meaningful, spanning open ocean, NTN coverage, and photonic backbone infrastructure, is unlikely to be resourced within any single national research programme. The bilateral research conditions that global investments in advanced connectivity have created are a practical foundation for the coordinated, multi-institution field programme this stage requires.

## V. Conclusion

The central argument of this paper is simple to state but difficult to execute. The five principal connectivity domains, optical fibre, free-space optical and RF wireless, non-terrestrial networks, maritime, and underwater optical and acoustic communications, have each developed substantial research literatures, mature channel models, and domain-specific standards. The connections between them have received far less attention, and the challenges that sit at those connections- compound channel statistics at medium boundaries, protocol behaviour across latency ranges spanning microseconds to tens of seconds, and orchestration across domains with fundamentally different physical dynamics- have no systematic treatment anywhere in the existing literature.

The CHAIN framework proposed here does not solve these problems. It names them precisely, locates them in relation to each other, and argues that they must be addressed together rather than one domain at a time. The closest existing work, SAGIN at three domains and SAGSIN at the sea surface, has established the theoretical case for multi-domain integration without producing the experimental and standards infrastructure to realise it. CHAIN extends that trajectory to all five domains, including the subsurface, adds the all-photonic network backbone as the unifying physical infrastructure, and introduces AI-driven orchestration as the cross-domain control layer, at a moment when those architectural choices are still open rather than inherited from prior frameworks.

The five open problems identified in this work have a dependency structure that determines the research sequence. Cross-domain channel characterisation must precede protocol design, which must precede serious orchestration work, which must precede the standards engagement that makes deployment possible. The near-term priority is establishing the measurement infrastructure, a co-located multi-domain facility where the statistics of cross-medium propagation paths can be observed directly, because without it the compound channel models that underpin everything else cannot be validated.

The convergence advantages should yield improved resilience from path diversity, extended coverage into the ocean volume, efficiency gains from the photonic backbone, and performance optimisation through workload-to-medium matching, are supported by two-domain precedent but unproven at five-domain scale. Proving them is the medium and long-term task, and it is one that requires the kind of international collaborative research infrastructure that no single national programme is positioned to build alone. A national facility encompassing the wide variety of domains is key to enabling this level of testing on a global scale.